\documentstyle[12pt]{article}
\title  {Mode Solutions of the Klein-Gordon equation \\
in warped spacetimes}

\medskip

\author { Philippe Droz-Vincent\\[2mm]
Laboratoire de Gravitation et Cosmologie Relativistes\\
C.N.R.S./ ER 2057, Universit\'e Pierre et Marie Curie\\
Tour 22-12 , boite courrier 142,\\
 4 place Jussieu 75252 Paris Cedex 05, France}
\date {30.03.00}     
\newcommand{\mun}{ {\mu  \nu} }
\newcommand{\albe}{ {\alpha  \beta} }

\newcommand{\ij}{ {ij} }
  
\newcommand{\AB}{ {AB} }  

\newcommand{\beq}{\begin{equation}}
\newcommand{\eeq}{\end{equation}}
\newcommand{\alp}{\alpha}             
\newcommand{\gam}{\gamma}
\newcommand{\Gam}{\Gamma}
\newcommand {\half}{ {1 \over 2}} 
\newcommand {\noi}{\noindent}

\newcommand {\lam}{\lambda}

\newcommand  {\disp}{\displaystyle}
\newcommand {\nabtil}{  {\widetilde \nabla}   }
\newcommand{\del}{\delta}     
\newcommand {\Del}{\Delta} 
\newcommand {\Sig}{\Sigma}

\newcommand {\fhat}{\widehat f} 
\newcommand \hhat {\widehat h}
\newcommand {\nab}{\nabla}
\newcommand{\nabo}{\nabla _1}
\newcommand{\alptil}{  {\widetilde \alpha}   }
\newcommand{\eron}{{\cal E}}
\newcommand{\sron}{{\cal S}}
\newcommand{\hron}{{\cal H}} 
\newtheorem{prop}{Proposition}
\newtheorem{theo}{Theorem}     
 \newtheorem{lem}{Lemma}
\begin{document}
\maketitle
\abstract{
In order to reduce the Klein-Gordon equation (with minimal coupling), we 
introduce a  generalization of the so-called  "mode solutions" that are well
 known in the special case of  a  Robertson-Walker universe.
  After separation of the variables, we end up with  a partial differential 
equation in lower dimension.  A reduced version of the Gordon
 current arises and is conserved. When the first factor-manifold is Lorentzian, 
 distinct modes appear as orthogonal in the sense of the sesquilinear form 
associated with the Gordon current.  Moreover,  a sesquilinear form is defined
 on the space of solutions to the {\em reduced} equation.
Extension of this picture to curvature  coupling is possible when the second
factor manifold is of constant scalar curvature.}

\section {Introduction}

A warped spacetime \cite{oneil}, \cite {desz}, \cite {erlich} is a product 
manifold $V_1 \times V_2$ 
 endowed with (omitting the canonical projections) a  metric
\footnote{The minus sign  %in  equation  (\ref{ds2}) 
 is dictated by our signature  $+- \dots  -$.   See next Section.}          
 \beq    (g) =  \alp \oplus (-S) \gam          \label{oplus}    \eeq
where $\alp$ and $\gam$ respectively are  metric tensors on
 $V_1$ and $V_2$.
Either $\alp $ or $\gam $ is Lorentzian, so that $(g)$ is normal hyperbolic.
$S$  is a positive  function  on $V_1$, it is convenient to set
$S = {\rm e} ^{2i \Theta} $.

\noi The most simple  example  of warped spacetime  is   probably given by the 
 Friedman-Robinson-Walker (FRW)  universe.
But the warped structure accomodates a large number of  metrics of 
physical interest in General Relativity.

\noi  In FRW, the wave  equation has been studied in details for many decades,
 in relation with  the old prblem of defining its positive-frequency solutions.
 To this end, the general solution is usually developed over the so-called
 {\it mode solutions}, so the Klein-Gordon equation gets reduced to an 
ordinary differential equation involving  the time variable only.

\noi In fact the property of being spatially homogeneous, exhibited by FRW 
universes, plays no role in this reduction, which permits to follow the same 
line in {\em generalized FRW} spacetimes  \cite{LMP} where the spatial sections
have  not necessarily a constant curvature.
The  geometrical properties of these spacetimes have been 
 systematically investigated by M. S\'anchez  \cite{sanch}.  

\noi  Warped structures in general have been widely  studied in the 
literature,  the main issue being of course to establish  relations 
between symmetry and curvature properties of the total spacetime and that of
 its factor manifolds. 

\medskip
\noi In this article we are concerned with interesting features exhibited by 
 the  Klein Gordon equation in a warped spacetime of arbitrary dimension and
 type (with some preference however for the case where $ V_1 $ is Lorentzian).
Assuming the  minimal coupling we write
\beq ( \nab  ^2   +  m^2 )  \Psi = 0            \label{1}   \eeq
for a complex $c$-number-valued wave  function. 

\noi In spite of well-known limitations, a particle interpretation 
of (\ref{1}) remains of interest. In the operator approach to quantum field 
theory in curved spacetime, one-particle concepts play at least the role of 
useful tools.
For instance, the kernel which projects any solution onto a
 positive-frequency  subspace and determines a definition of the vacuum, is 
a solution of the KG equation (often referred to as the two-point Wightman
function).

Our goal is to show that most results obtained up to now in the context of  
generalized FRW spacetimes can be systematically extended to any kind of 
warped spacetime. 

\noi
We shall take advantage of the  warped-product structure in order to carry 
out a systematic reduction of (\ref{1}), and shall  end up  with an
 equation to be solved for a reduced wave function which depends only on the 
 coordinates running in  the first  factor manifold  $V_1$.

\noi
The principle of this procedure consists in developing the general solution 
of equation  (\ref{1})  over special ones that generalize the mode solutions 
arising in the customary FRW framework.

\noi The possibility of reducing the KG equation with help of generalized 
modes can be tracked back to a remarkable feature of classical motion in 
warped spacetimes. That point is briefly exposed in Section 2, in parallel
 to a study of the differential operator which describes the quantum
 motion.         Generalized mode solutions are introduced in Section 3, 
where the separation of variables is  carried out.

\medskip  
\noi
Together with the KG equation, we analyse  the sesquilinear form 
defined on its solutions.  Our  motivation  is the fact that,
  for complex (resp. real) solutions 
sesquilinear (resp. bilinear) forms are  fundamental with respect to
  quantum mechanics;
they provide a framework for the  construction of a "complex structure 
positive operator" ensuring the splitting of any solution into positive and 
negative-frequency parts  \cite{seglich} , \cite{moreno}.

\noi We shall demonstrate in Sections  4-5  that  the sesquilinear form 
defined for solutions to equation (\ref{1})
(by  conservation of the Gordon current)  undergoes some kind of reduction.
Indeed, under very general assumptions, the space of complex solutions to the 
{\em  reduced equation} can be  in turn endowed with a sesquilinear form of 
its own.
  Actually,  a conserved  vector-density defined on the first factor manifold
is  associated with the reduced wave equation; we display 
the  relationship of this object with the usual Gordon current.

\noi The possibility of 
extending this study to nonminimal couplings is discussed in Section 6.

\medskip
\noi    Although we have in mind possible applications to quantum mechanics, 
this article is written from the viewpoint of differential geometry;
all functions and tensors are supposed to be {\sl smooth}, that is $C^\infty$.

\noi Moreover all manifolds considered here are implicitly assumed to be 
{\sl connected}.

\subsection{Notation}

In the product 
\noi      $ (V)  =  V_1  \times  V_2 $,  the factors   $V_1, \  V_2 $ 
     respectively have dimensions  $p, \  q$. 
 \noi We define  Type I (resp. Type II)   by this property that 
 $(V_1, \alp ) $  (resp.    $(V_2 ,  \gam ) $ )  is   Lorentzian.           
 
\noi  Throughout this paper we use coordinate charts  adapted to the warped 
structure. The metric takes on the  orthogonal form

\beq ds^2 =  \alp _ \AB     \   dx ^A  dx^B  -
  S(x^C)  \gam (x^k) _ \ij   \     dx^i dx^j    \label{ds2}   \eeq
where $  \disp A, B, C  \in I_1,  \quad   i,j,k  \in I_2 
\quad     I_1  \cap I_2 = \oslash  $
and  $ I_1 \cup I_2  $ covers the whole set of integers $0,1,2,...., p+q-1$.

\noi [For instance in 4 dimensions:

\noi  $I_1 \ni 0,  \quad I_2  \ni  1,2,3 \quad   $ for  FRW spacetimes,
        and we can take        $I_1 \ni 0, 3   \quad I_2  \ni  1, 2  \quad $ 
 for spherically symmetric universes]

\noi    The minus sign in equation  (\ref{ds2})  is dictated by our signature
 $+- \dots  -$.                                                              
   Type I   necessarily  corresponds to  $\gam$ positive definite.
In contrast, Type II implies  having the quadratic form  $\alp$   elliptic but 
{\it negative}  definite.

\noi      $ g_\AB  =  \alp _\AB$              and 
      $ g_ \ij    =  - S  \gam _  \ij    $.

\noi [Example: In FRW spacetimes, , $p = 1 , \quad q = 3$ and we have
  $A, B, C = 0$  and $i, j, k = 1, 2, 3$].

It is clear that 
    $$   g_\mun =  
    \left(      \begin{array}{cc}
       \alp _\AB     &    0               \\
        0             &     -S \gam _ \ij   
\end{array}           \right)    
\quad    ,  \qquad \quad 
 g^\mun =  \left(   \begin{array}{cc}
    \alp ^\AB     &    0    \\
  0           &    -S^{-1} \gam ^\ij  
\end{array}          \right)   $$

\noi   where   $\disp    g^\AB   g _{BC} = \del  ^A _C   ,  \qquad    
        g^\ij    g_ {jk} = \del ^i _k  $
Notice that 
$$     g^\AB =   \alp ^\AB   $$
if we define $\alp ^\AB $ (resp. $\gam  ^\ij $)  
as the  contravariant tensor inverting  $\alp _\AB $ (resp. $\gam _\ij$), 
  that is  
$  \alp ^\AB    \alp  _ {BC}  = \del  ^A _ C \quad, \qquad 
 \gam ^\AB    \gam  _ {BC}  = \del  ^A _ C     $.
We obviously can write 
  $  g _\AB =   \alp  _ \AB   $.

\noi   Notice that 
$$  g^\ij  =  (-S) ^{-1} \gam ^\ij                               $$

\noi  Caution that the  Types I, II defined above must not be confused with the 
Classes $A, B$  introduced by Carot and da Costa for  four-dimensional
  warped spacetimes \cite {carot}.
For $p+q=4$, intersection of Classes $ A_1, \quad A_2$ and $B$ with both
 Types give rise to six  possibilities.

\noi In references   \cite{carot} \cite{sanch} each class is combined with
 the Types by chosing a sign $\pm$ in the generic form of the metric.
The contact with the   notation  of Ref. \cite{carot}  can be made as follows:  
$$   \alp = - h_1 ,    \qquad   \gam = h_ 2                    $$
where $h_1$ and $h_2$ are the metrics assigned to the factor manifolds
 in Ref. \cite{carot}.
Our quadratic form $ds^2 = g_\albe  dx^\alp   dx ^\beta$ 
and their have opposite signs.

\medskip
\noi  Geometric objects corresponding to  
 $ (V_1 , \alp ) , \quad (V_2 , \gam )$  are  affected by the index $1, 2$
respectively. This label will be put on the left for connexions,  curvature
tensors and their contractions.

\medskip
\noi Remark

\noi  It is a trivial observation that the Minkowski  metric is (globally) 
decomposable in several ways. Similarly, it may happen that a given spacetime 
can be considered as warped in several ways, so that, for instance, it may be 
of Class $A$ for one structure whilst it is of Class $B $  for another one 
\cite{carot}. But here, we consider only one structure at one time, so the 
six possibilities mentioned above are mutually exclusive.

\medskip
$ \eta = \sqrt{|g|}  \varepsilon $ where $\varepsilon $  is the Levi-Civitta 
tensor.
\noi   Useful determinants are as follows: $ \quad   $    Setting 
$$ g = \det  \   g_\albe   \qquad   \gam = \det  \   \gam _ \ij  
    \qquad    \alp  =  \det \   \alp _\AB  \equiv \det  \   g _ \AB          $$ 
we have  
    $$    \det  \   g_\ij    =  (-S)^{q}  \     \det \    \gam _ \ij $$
and therefore
 $ \disp    g =  \alp    \     (-S)^{q}  \    \gam                        $
$ g = \det  \   g_\AB  \   (-S)^{q} \    \det  \   \gam _ \ij    $   thus  
$  g =   \alp   \   (-S)^{q}  \     \gam      $. We shall rather use
\beq \sqrt{ |g|} =  
\sqrt {|\alp |} \  S^{q /2}  \    \sqrt {|\gam |}          \label{det}  \eeq

\noi  Volume elements: $ \quad $   Setting
$$ d_1 ^p \    x  =  \bigwedge   dx^A ,     \qquad
 d_2 ^q \    x =
 \bigwedge   dx^j ,  \    \qquad  A \in I_1 ,   \qquad j \in I_2            $$
we have  $\disp  \qquad  d^{p+q} x = d_1 ^p \     x   \wedge  d_2 ^q \   x  $.

\noi   The volume elements of  $  (V_1 , \alp )$ and  $ (V_2 , \gam )$
 respectively  are
$ \sqrt {|\alp |} \    d_1 ^p x $  and     $ \sqrt {|\gam |} \    d_2 ^q x $,
whereas naturally  the volume form of  $(V)$ is 
            $ \sqrt {|g|} \    d ^{p+q} \      x  $.

\medskip
\noi
For Type I,  it is convenient to take $x^A $ running from $0$ to $p-1$ and to
factorize out the time coordinate by setting 
\beq  \omega =
   d x^1 \wedge  \dots  \wedge dx^{p-1} \label{defomeg}      \eeq
so that  $   \disp    d_1 ^p \    x  =    d x^0  \wedge \omega    $.

\noi With this convention, we have
\beq     \omega = {1 \over   (p-1)!}   \
\varepsilon  _{0 B_1 \dots  B_{p-1} }   \
dx^ {B_1} \wedge \dots   \wedge dx^ {B_{p-1} }         \label{defomeg2}  \eeq

\bigskip

\section {Classical and Quantum Motion in Warped Spacetime}

\subsection{Geodesic motion}
\noi
An equation like (\ref{1}) can be thought of as describing the quantum 
motion of 
a test particle, in the approximation where possible particle creation is 
neglected.         It is in 
order to point out that, in any warped spacetime,  the classical motion of a 
free particle (geodesic  motion)  already enjoys an interesting property which
 directly stems from the warping.

\noi Indeed the equations of motion of a test particle in $V$ are canonically
 generated by   a  "Hamiltonian  function" 
\cite{pomme}
  $$ G (x, p)  = \half g^\albe   \  p_\alp  p_ \beta   
  = \half (  \alp ^\AB p_A p_B - S^{-1}  \gam ^\ij \    p_i p_j )         $$
which is  a scalar  in  the cotangent bundle   $ T _ * (V)$.  According to 
the canonical  symplectic form of this bundle, we have the usual Poisson 
brackets
 $  \disp   
\{  x^\alp , p _\beta  \}   =  \del  ^\alp _ \beta    $ , etc.
Constants of the motion are characterized by a vanishing Poisson bracket
 with $G$.       It is easy to verify that

\begin{prop}
In any warped spacetime, with the metric written like in (\ref{oplus}),
 geodesic motion admits  the first integral
\beq     2 K =      \gam ^\ij  \   p_i p_j           \label{calk}  \eeq
where  $p_\alp $ are the momenta.
\end{prop}

 [  Proof:        We see that 
$   \{ K ,   \alp ^ \AB \    p_A  p_B  \}   $ vanishes. 
This is obvious since $K$  only depends on
$ x^C , p_D$.   Then we observe that $ \{ p_j  , x^A  \} = 0$. It follows that 
$  \{  K , S   \} = 0$,  so finally
$ \disp   \{  g ^\albe  p_\alp  p_\beta ,  K    \}   = 0     $ and $K$
is a constant of the motion.]
Indeed we derive with help of the standard Poisson brackets that
$ \{ G,  K \} = 0$.

\noi        For Type I, the quantity $ K/m $  somehow generalizes
 the kinetic energy, although its conservation is ensured even if
 $(V_2 , \gam )$ fails to admit a group of translations.
For instance, 
when $V $ is simply  ${\bf R} \times {\bf R}^3$ warped with some time-depending 
scale factor, the conservation of  $K$ can also  be  derived from  the 
existence of a translation group in  $ {\bf R}^3$. And in this case,
 $K/m$  is just the kinetic energy in the usual sense.
But the point is that this  property survives when  $ {\bf R}^3$  is replaced 
by any other  three-dimensional manifold.

\bigskip
\subsection{Quantum  motion}

\noi  We assume minimal coupling, so we write the KG equation as  (\ref{1})
where   
$ \nab ^2  \Psi   = g ^\albe  \nab _ \alp \nab _ \beta \Psi  $.     But we 
shall use the well-known formula
\beq       \nab ^2 \Psi =   {1 \over  \sqrt |g| } 
 \partial \mu  (  \sqrt {|g|}  \     g^\mun  \partial _ \nu  \Psi )  
\label{rac}         \eeq
By formal analogy with (\ref{calk}) it is natural to consider  that the quantum
 mechanical analog of $K$  is   
$ K_{\rm quant}    =  - \half  \Delta _2$  where 
$\Delta _2$  is the Laplace-Beltrami operator in  $ (V_2 ,  \gam ) $.
Indeed we have in obvious notations 
$\Delta_2 = \gam ^\ij (\nab_2)_i  (\nab_2)_j $ acting on scalars.
 But we rather use the formula
$$ \Delta _2  \Psi =  {1 \over \sqrt {|\gam |}}\partial_i (\sqrt {|\gam |} 
\gam ^\ij \partial _j   \Psi)                      $$

\noi   Developing  formula (\ref{rac}) we get 
$$   \sqrt{|g|}  \   \nab ^2   \   \Psi = 
\partial _A (  \sqrt {|g|}   g  ^\AB \partial _B  \Psi) +
\partial _i (  \sqrt {|g|}   g  ^\ij \partial _j  \Psi)     $$
but  $  g ^\AB =  \alp ^\AB $     and   
$  g^\ij  =  -S ^{-1} \gam ^\ij  $   thus
$$   \sqrt{|g|}   \nab ^2 \Psi =
\partial _A (  \sqrt {|g|}   g  ^\AB \partial _B  \Psi)    -
\partial _i ( \sqrt{|g|} S^{-1} \gam ^\ij   \partial _j  \Psi )  $$
where $ \partial _ k S = 0$.     But  in view of (\ref{det}) we have
$$   \sqrt{|g|}  \   \nab ^2   \   \Psi = 
\partial _A (  \sqrt {|g|}   g  ^\AB \partial _B  \Psi)
    -    S ^{-1}     S^{q /2}   \                     
\partial _i  (\sqrt {|\alp |} \sqrt {|\gam |}  \  \gam ^\ij  \  
\partial _j \Psi )    $$                                      
Since $\alp$ only depends on $x^A$ we obtain
$$   \sqrt{|g|}  \   \nab ^2   \   \Psi = 
\partial _A (  \sqrt {|g|}   g  ^\AB \partial _B  \Psi)
    -    S ^{-1}     S^{q /2}   \  \sqrt {|\alp |} \
\partial _i  ( \sqrt {|\gam |}  \  \gam ^\ij  \  
\partial _j \Psi )    $$
Again develop  $\sqrt {|g|} $ and remember that 
 $ \disp  \partial _ A  \gam = 0 $.
We get 
$$   \sqrt{|g|}  \   \nab ^2   \   \Psi =
\sqrt {|\gam |} \partial _A ( \sqrt {|\alp |} \      S^{q /2}  \
g^\AB \partial _B \Psi )    
  -    S^{-1}  S ^{q / 2} \   \sqrt {|\alp |} \  
\partial _i (\sqrt {|\gam |} \   \gam ^\ij  \ \partial _j   \Psi  $$
Again develop    $\sqrt {|g|} $ hence
$$   \nab ^2   \   \Psi  =
{1 \over    \sqrt {|\alp |} \     S ^{q / 2} }
\partial _A (    \sqrt {|\alp |} \     S ^{q / 2}  \ 
  g^\AB  \partial _B  \Psi )   -
{ S^{-1} \over   \sqrt {|\gam |} }  \ 
 \partial _i ( \sqrt {|\gam |} \   \gam ^\ij  \ \partial _j   \Psi  )     $$
It is convenient to define  
$$ D \Psi =
{1 \over   \sqrt {|\alp |}  }  S ^{- q / 2}
\partial _A (    \sqrt {|\alp |} \     S ^{q / 2}  \ 
  g^\AB  \partial _B  \Psi )                                              $$
irrespective  of whether  $\Psi$ is a solution to (\ref{1})  or not.

\noi
Indeed  the second order differential operator  $D$  only affects quantities
depending on the $x^A$ variables. 
So we can write
$$    \nab ^2   \   \Psi  =  D \Psi   - 
S^{-1} \    {1\over   \sqrt {|\gam |} }  \   \partial _i
(     \sqrt {|\gam |} \gam ^\ij \partial _j  \Psi )  $$
In other words we have the identity
\beq    \nab ^2   \   \Psi  =  D \Psi   -
S^{-1} \      \Delta_2   \   \Psi      \label{reduc}             \eeq
where    $ \Delta _2  $  is the  $q$-dimensional Laplace-Beltrami 
operator, associated with the manifold  $ (V_2 ,  \gam )$.
As an operator extended to functions  on $V$, it does not 
affect the quantities depending on $x^A$ only.

It is clear  that   $\Del _2$  commutes with   $D$, 
 because these operators  act on separate sets of variables.
 As $S$ does not depend on the $x^j$ coordinates, 
it is clear that  $\Del _2$ (or equivalently $K_{\rm quant}$)  commutes
 with $\nab ^2 $.
 In the classical limit
 ($\hbar \rightarrow 0$) this property reduces to the conservation of $K$.

\medskip
\noi
Let us re-arrange $D$ in order to simplify the expression of $\nab ^2 \Psi $.

\noi   Let us provisionally use coordinates where $ |\alp |= 1$. We obtain
$$ D \Psi =  S ^{- q / 2}   \partial _A
 ( S ^{q / 2}  g^\AB \partial _B \Psi ) $$
$$D  \Psi =     \partial _A   (  g^\AB \partial _B \Psi )
+   {q  \over 2} (\partial _A  \log S )  g^\AB \partial _B \Psi          $$
But $g^\AB  =  \alp ^\AB $ and $|\alp | = 1$, thus
\beq D \Psi = \Delta _1  \Psi + {q \over 2} \alp ^\AB  (\partial _A \log S)
\partial _B  \Psi                 \label{depsi}                      \eeq
which is valid  in all coordinates  and for arbitrary $\Psi$. 
This  expression, where the coordinates $x^j$ are ignorable, 
 is to be inserted into equation (\ref{reduc}).

\bigskip

\section{Mode Solutions, Product Solutions}

 Since $ [\Del _2, \nab ^2 ] $  vanishes, it is clear that  some solutions of 
the KG equation are also eigenstates of  $  \Delta _2  $.
If   $\Phi $ is such a solution we have 
\beq         \Delta _2  \Phi  =  - \lambda  \Phi    \label{lapl}    \eeq
for some constant number $\lam  \in   {\rm Spec.} (V_2) $.
We shall generalize the terminology which is commonly used when $V$ is FRW 
universe and shall call  $\Phi $ a {\sl Mode Solution} to the KG equation.
For a  solution in mode $\lambda$,  the KG equation  reduces to
\beq (D +  \lam S^{-1}  +  m^2 )   \Phi   = 0       \label{mod}           \eeq
where the differential operator $D$  acting on $\Phi$ affects the $x^A$'s 
only.

\noi       Finally, $x^A$ and $x^j$  are respectively ignorable in 
equations (\ref{mod}) and (\ref{lapl}), which realizes separation of the 
variables. 

\noi Leaving aside the solving of      (\ref{lapl}),                           
the original KG equation in    $(V)$ has been reduced to a 
 (linear) partial differential equation in $p$ dimensions.

\medskip
\noi From now on, we look for solutions to     (\ref{1}) in the form of
 a superposition
of various mode solutions corresponding to all possible values taken by $\lam$ 
in the spectrum of $V_2$.

\medskip    \noi
 In the special case where    $p = 1$,  equation  (\ref{mod})
  is an ordinary 2nd order equation and its solutions form a two-dimensional 
vector space.

Otherwize, the space of solutions still has infinitely many dimensions.

\bigskip
\noi Some special solutions of the wave equation (\ref{1}) have the form of 
a {\sl product} of functions compatible with the product  structure 
of spacetime, namely
 \beq \Phi = f (x^A) \  F (x^k)      \label{produi}          \eeq
We observe that

\begin{prop}
Any product solution $fF$ to the KG equation is a mode solution,
and   $F$  is eigenfunction of  $\Del _2$ .
\end{prop}
\noi   Proof.

$$ \nab ^2 \Phi =  (Df) F  -  S^{-1} \   f  \   {\Delta}_2  F   $$ 
$$ (\nab ^2  +  m^2 ) \Phi = F (D+m^2) f  
        -  S^{-1} \   f  \   {\Delta}_2  F   $$
According to KG equation this quantity vanishes.   
$$  F (D+m^2) f  
       =  S^{-1} \   f  \   {\Delta}_2  F   $$
  Discarding a trivial case, neither $f$ nor  $ F$ can identically vanish.
When $f$ and $F$ are not zero, we divide by $fF$  and multiply by $S$. We get 
$$  S  {  (D+m^2) f  \over  f}  
       =     {  {\Delta}_2  F  \over  F}          $$
The l.h.s. of this equation depends on $x^A$ only while the r.h.s. only 
depends on $x^k$. Both are thus necessarily constants, so there exists some 
$\lam $ such that
  \beq {\Delta}_2  F = -   \lam    F    \label{lapl2}     \eeq

\medskip
\noi        Let  $ \eron [\lam ] $ 
be the space of smooth functions on $V_2$ satisfying (\ref{lapl2}) for a 
given value of $\lam$.
For  $V_2$ compact,  $\Del _2$ has a discrete spectrum which is 
the infinite sequence 
$$ {\rm Spec} (V_2 , \gam ) = 
 \{ \lam _0 = 0, <  \lam _1 , \dots  < \lam _n \dots \}$$
In this case $\eron _n $ denotes  $\eron [\lam _n] $ and we know that 
its dimension  $d(n)$  is {\em finite} \cite{berg}.

\bigskip
\noi  
The converse of the previous  Proposition is not true, but

\begin{prop}
 In a warped spacetime of Type I with  compact $(V_2)$, 
 any mode solution corresponding to a given
 $\lam$ in $ {\rm Spec} (V_2,  \gam)$ is a  finite sum  of product solutions.
 \end{prop}
 Let $f_u (x^A)$ be
 the coefficients  of $\Phi $ in a development over a  basis
$ F_1 (x^j), \dots \dots  $.
\beq  \Phi = \sum_{u=1 } ^{d(n)}     f_u (x^A)  \    F_u (x^j)    
                                                  \label{basedev}    \eeq
\beq   (\nab ^2 + m^2) \Phi = \sum F_u (x^j) \
       [D + S^{-1} \lam_n  + m^ 2] \   f_u (x^A)      \label{prodmod}  \eeq  
This expression must vanish. Since $F_1 \dots  F_d $ form a basis,   it is  
 clear that each   $f_u$  must be a solution to the equation 
  \beq  (D + \lam _n  S^{-1} +  m^2)  f  = 0      \label{fmod}      \eeq
It follows that  each $f_u F _u$ is a product solution.

\medskip
\noi Let  $\sron [\lam ] $ be the space of smooth   functions on   $V_1$
satisfying to    (\ref{fmod}) for a given value of $\lam$.
                                                                     
\noi 
Except in the very special case where $p=1$, (\ref {fmod}) is a {\em partial} 
differential equation and  has infinitely many linearly independent 
solutions.

\medskip
\noi
Developping  $D$ with help of (\ref{depsi}) we   obtain after simplification
\beq  \Delta _1  f + {q \over 2} \alp ^ \AB (\partial _A \log S)
 \partial _ B  f    + (\lam S^{-1}   +    m^2 )  f    = 0  
                                                 \label {rreduc}          \eeq
which is a $p$-dimensional problem only, formulated in terms of the metric
 $\alp $.     The $x^j$ do not arise in this equation.

\medskip
\noi    To summarize, equation (\ref{1}) has been reduced to a pair of 
equations  involving separate sets of variables.  
These equations are (\ref{lapl2}) and (\ref{rreduc}). For Type I (resp. Type 
II) the Laplacian  in the former is elliptic (resp. hyperbolic) whereas  the 
partial differential operator in the latter is hyperbolic (resp. elliptic).
Notice that   (\ref{lapl2}) involves only the geometry of $V_2$ , whereas 
(\ref{rreduc}) involves not  only  the geometry of $ V_1 $
 but also the shape of the warping function $S$.  

\medskip
\noi
For Type I with  compact $V_2$,   we write  $\sron _n $
 for $\sron [\lam _n ]$.     The $n$th mode will be noted
$$  \hron  _n  =  \sron _n  \otimes  \eron _n$$
$$  \         $$
Now, a comparison of  (\ref{rreduc}) with  (\ref{1}) is in order:
Equation        (\ref{1}) simply involves the $p+q$-dimensional Laplacian, 
which implies that the Gordon current in $p+q$ dimensions is conservative.
In contrast,      (\ref{rreduc}) not only involves the $p$-dimensional 
Laplacian $\Del _1$ and an innocent multiplicative operator 
$   \lam  S^{-1}   + m^2   $,  but also first order partial differentiation.
As a result, the $p$-dimensional sesquilinear field constructed with a couple
   of solutions $f,  h $ to     (\ref{rreduc}), that is to say
\beq         I^A  =       -i (f^*  \   { \nabla }_1  ^A  \   h 
  - h \   {\nabla }_1  ^A   \   f^* )              \label{IA}  \eeq
{\em fails to be} divergence-free.

\noi For the same reason, the second order linear differential operator
 involved in (\ref{rreduc})   {\em  is not \/}
  symmetric with respect to the scalar  product
        $$     <f, h>_1   = \int _{V_1} f^*  h  \   \sqrt{|\alp|} \
        d_1 ^p \    x                              $$
   defined with help of  the $p$-dimensional volume element
 $\sqrt {|\alp |} \   d_1 ^p \    x $  determined by the metric  $\alp $ in 
$V_1$.

\noi This situation is  more specially unpleasant in Type I spacetimes,  where
 (\ref{rreduc}) includes the dynamical aspects of the original 
equation KG equation.
In this case,   a scalar product for any couple of solutions to 
 (\ref{rreduc})  would be of interest for  quantum  mechanics.
So,  it is desirable to construct some conserved $p$-dimensional current, 
sesquilinear with respect to the couple $f, h$.
 But the presence of  $\partial f$ in (\ref{rreduc}) is an obstacle.

\medskip
\noi  Before we focus our attention on Type I, 
let us first replace  the vector field of  formula   (\ref{IA})
by a better candidate in order to make up a conservation law.

[\noi
N.B. Statements about the divergence of a vector necessarily refer to a 
metric.
In contradistinction, the divergence of a vector-{\em density} is intrinsic.]

\noi  The  difficulty associated with the presence of first derivatives in 
(\ref{rreduc}) can be circumvented by two manners.
 \begin{itemize} 
\item   Use another  metric (conformal to  $\alp _\AB$)
  on the manifold  $V_1$ and re-write   (\ref{rreduc}) in terms of it.
This procedure amounts to consider another differential operator that is
 symmetric in the sense of this new metric.  
\item  Keep the metric of $V_1$ unaltered, but make a conformal change
 of function, say
$ f = S ^r  \fhat $.
\end{itemize}

\section{Conserved Currents} 
\subsection{First Method}

\noi  
Let us consider  in  $V_1$   a new metric $ \alptil _\AB$ such 
that 
$$ \alp  _\AB    =  U (x^C)    \alptil _\AB      $$  for some  conformal 
factor   $  U (x^C) $ which must be suitably chosen.   It is clear that 
$  \alp ^\AB  =  U^{-1}   \alptil ^\AB $, if  we call  $\alptil ^\AB $  the 
contravariant tensor inverting $\alp _ \AB $.
We set  $ \disp  \nabtil _1 ^A = \alptil ^\AB  \partial  _B $.

\noi The determinants are related by
\beq \det \alp _\AB  =  U ^{p} \   \det ( \alptil _ \AB ) \label{deter}  \eeq
In view of this formula,  we find
$$    \Delta _1  f =    U^{-1}  \    {\widetilde \Delta } _ 1    f   +
  ( \partial _A  \log   U ^ {p /2  -1} )  
 \   \alp ^\AB    \partial _B   f                                         $$
This is to be inserted into $D f$ which is given by (\ref{depsi}). We get
\beq    Df =      U^{-1}  \    {\widetilde \Delta } _ 1    f       +
  ( \partial _A  \log   U ^ {p /2  -1} )  
 \   \alp ^\AB    \partial _B   f         
+  {q \over 2} (\partial _A \log S )  \  \alp ^\AB  \partial _B  f   
                                                       \label {Df}      \eeq
                             
\noi The first derivatives of $f$ are eliminated from  $Df$  provided that
$$   U ^ {p /2  -1} =  {\rm const.}    \    S ^{-q / 2 }                   $$
which is possible (for nontrivial $S$) under the condition that  $ p \not=  2$.
 In this case it is sufficient to take 
$  \disp    U ^{p  - 2}  =    S ^{- q}         $,   which leads to
\beq  U   =    S  ^{q  \over   2-p }             \label{UdeS}         \eeq
Thus  (\ref{deter})  entails
\beq \sqrt {|\alp |} = S^{pq \over 4 - 2p} \sqrt{|\alptil |} 
                                                     \label{sqalp}      \eeq
With the  choice (\ref{UdeS}) we simply have
$$ Df =      U^{-1}  \    {\widetilde \Delta } _ 1    f             $$
to be inserted into equation (\ref{fmod})  for  mode $\lam$.

 We end up with an equation of the form
\beq  {\widetilde \Delta } _ 1    f  + Q  f   = 0       \label{eigen}   \eeq
where 
$ \disp   Q  =    U  ( \lambda    S ^{-1}   + m^2  )                  $
but  equation   (\ref{UdeS}) implies   that
$$ Q =    S ^{q \over 2-p} \    ( \lam  S^{-1}   +    m^2 )   $$
Notice that 
 $   {\widetilde \Delta } _ 1    + Q $   in  equation  (\ref{eigen}) 
          is symmetric with respect to 
the scalar product      $ \disp   <\widetilde {f,h} >_ 1  $,   defined with 
help of the $p$-dimensional volume  
$ \sqrt {|\alptil | }   \        d_1 ^p \     x $.
 Moreover,  the $p$-dimensional  current 
\beq  \disp   {\widetilde I} ^A (f,h) =
 -i (f^*  \   {\widetilde \nabla }_1 ^A  \   h 
  - h \   {\widetilde \nabla }_1 ^A   \   f^* )            \label{pGord}  \eeq
is divergence-free in $(V_1 ,   {\widetilde \alp})$
 for any couple of solutions $f, h$ to the same equation 
(\ref{eigen}). We mean the same $\lam$ in (\ref{eigen}) for $f$ and $h$ and
  of course  
 $ {\widetilde \nabla }_1 ^A =  \alptil ^\AB  \partial _B $ .
In other words we have 
$$  \partial  _A  (   \sqrt { | \alptil |} \   {\widetilde I} ^A)  =  0     $$
Remark: 
When $p=1$ then ${ \widetilde I}^A$ has a single component  only, so the
 above conservation law reduces to the well-known constancy of the Wronskian.

\medskip
\noi   
So, provided $p \not= 2 $,  it is  possible to eliminate 
$ \alp ^\AB  \partial _B \log S $     from  (\ref{Df})  by chosing $U$
as in (\ref{UdeS}).

\noi  It is noteworthy that in four-dimensional warped spacetime,  we 
precisely have   $p = 2 $ in the case of Class $B$. Cases with $p=2$ will be 
handled by  the second method  (next section).

\subsection{Second  Method}

\noi
Let us try a conformal change of function involving a suitable power of $S$.
So we introduce  $\fhat$ by setting
  $  f = S^r   \fhat  $, for some $r$ to be determined below. 
Our goal is to eliminate  $ \partial  \fhat$.

\noi Define in any dimension the Laplace-Beltrami operator on scalars
  $\Delta u =  \nab \cdot \nab u $. Well known that 
$$\Del (uv) = (\Del u ) v + u \Del v + 2  \nab u  \cdot   \nab v          $$
We apply this formula in manifold $V_1$, where the Laplacian operator is 
  $\Del _1$, to the product  $ S^r \fhat $.
In the present case
$$ \Delta _1  f =  (\Del _1  S^r ) \   \fhat  + S^r  \Delta _1 \   \fhat
+ 2 \alp ^ \AB (\partial _A S^r ) \   \partial _B \fhat $$
But (\ref{depsi}) tells that 
$$ D f =        \Delta _1  f + 
 {q \over 2} \alp ^\AB  (\partial _A \log S ) \   \partial _B f     $$
where  $\disp   \partial _ B f =   (\partial _B S^r ) \   \fhat   
 +  S^r   \partial _B  \fhat    $. Thus
$$  Df =  \Delta _1  f      + 
 {q \over 2}  \alp ^\AB (\partial _A  \log S ) (\partial _B S ^r ) \   \fhat
  +  {q \over 2}  \alp ^\AB    (\partial _A  \log S )   S^r   \partial _B  
\fhat    $$
First order derivatives  of $\fhat$ get cancelled from (\ref{rreduc})
 provided that    $ \disp   2 \partial _A  S^r 
  + {q\over 2} S ^r \partial _A  \log S  = 0    $.   Thus
\beq   r = - {q \over 4}       \label{rdeq}   \eeq
We are left with
$$  D f =  (\Del _ 1  S ^r ) \   \fhat  + S ^r \    \Del _1    \fhat 
+ {q\over 2}  \alp ^\AB (\partial _A  \log S) (\partial _ B S ^ r )
\   \fhat                          $$
$$  S ^{-r} D f =  S ^{-r}  (\Del _ 1  S ^r )\    \fhat    +
  \Del _1    \fhat    +   {qr \over 2}  \alp ^\AB   (\partial _A  \log S) 
(\partial _ B  \log S ) \  \fhat                                      $$ 
In view of  (\ref{rdeq}) we conclude that
\beq  S ^{q  / 4} D f =   \Del _1    \fhat   +  
 (S ^{q / 4}  \  \Del _ 1  S ^{-q/4} )\    \fhat    
-   {q^2 \over 8}  \alp ^\AB   (\partial _A  \log S) 
(\partial _ B  \log S ) \  \fhat                   \label{clef}         \eeq 
Equation (\ref{mod}) yields 
$ \disp   (D +  \lam S^{-1}  + m^2 )  \   f = 0  $. 
Multiply  by $ S ^{q/4}$, and use  $ \disp f = S ^{-q/4} \fhat $.
 We get
$$ S ^{q / 4} D f +  (\lam S^{-1}  + m^2 ) \   \fhat = 0            $$
where the  first term is to be developed in terms of $\fhat$ as in 
(\ref{clef})

\noi    We end up with a reduced equation of the form
$ \disp    \Del _1    \fhat    +   \Xi  \fhat   = 0  $.

\noi If $\fhat $ and $\hhat $ are two solutions to this equation (for the
 same $\lam$) they correspond to $f$ and $h$ through the formulas
\beq  f = S^r \fhat , \qquad \   h =  S^r \hhat        \label{fdefhat}      \eeq
and    the  $p$-dimensional current  
\beq     J^A (f, h)  =   I ^A (\fhat , \hhat )
=  -i ( \fhat ^* \nabo ^A \hhat 
 -  \hhat  \nabo ^A \fhat ^* )          \label{Gordhat}  \eeq
 is conserved
    $$ \partial _A   {\sqrt |\alp | }  J^A  =  0                           $$

\bigskip   \noi
In order to compute $J^A$ we use  (\ref{fdefhat}) and write 
\beq \nabo ^A \fhat ^* =  \alp ^\AB S^{-r} \partial _B \   f ^*
+ \alp ^\AB  ( \partial _B S ^{-r} ) \    f^*           \label{nabofhat} \eeq
\beq \nabo ^A \hhat  =  \alp ^\AB S^{-r} \partial _B \   h 
+ \alp ^\AB  ( \partial _B S ^{-r} ) \    h           \label{nabohhat} \eeq
Inserting  into (\ref{Gordhat})  yields
$$ i J^A  = 
S^{-r} f ^*   \  [    \alp ^\AB S^{-r} \partial _B \   h 
+ \alp ^\AB  ( \partial _B S ^{-r} ) \    h ]        
-   S^{-r} h   \    [ \alp ^\AB S^{-r} \partial _B \   f ^*
+ \alp ^\AB  ( \partial _B S ^{-r} ) \    f^*  ]                 $$
%So we get   
%$$ i J^A =
%\fhat ^*  S^{-r}    \nabo ^A  h  -  \hhat  S^{-r}    \nabo ^A   f^*  +
%\fhat ^* \    (\nabo ^A S^{-r}) \   h
%           - \hhat   (\nabo ^A S^{-r}) \       f^*         $$
After simplification we are left with
       \beq i J^A =
       \fhat ^*  S^{-r}    \nabo ^A  h  -  \hhat  S^{-r}    \nabo ^A   f^* 
               \label{iJA}       \eeq
After cancellation of two terms  we obtain 
            $$ i J^A  = 
S^{-2r}     (f ^*   \      \alp ^\AB  \partial _B \   h 
-    h   \     \alp ^\AB   \partial _B \   f ^*  )               $$
Remember that   $ -2r =  \half  q $. So finally
            \beq i J^A  = 
S^{q/2}     (f ^*   \      \alp ^\AB  \partial _B \   h 
-    h   \     \alp ^\AB   \partial _B \   f ^*  )    \label{iJAbis}    \eeq
\beq     J^A =   S^{q/2} I^A               \label{JdeI}      \eeq

\medskip
 \noi  Now, under the assumption that (\ref{UdeS})   and  (\ref{rdeq})
   hold true we can assert

\begin{prop} 
If  $p \not= 2$ then the vector fields   
$ {\widetilde I} $ on  $(V_1 , \alptil )$  and  $J$ on  $(V_1 , \alp )$
correspond to the same (conserved) vector-density, in other words 
$$  \sqrt{|\alp |} \   J^\alp =  \sqrt{|\alptil |} \    {\widetilde I}^\alp
         $$
   \end{prop}

Proof

\noi   Start from (\ref{Gordhat}), that is
 $$ i J^A =  \fhat ^*   \nab _1 ^A \hhat  - \hhat   \nab _1 ^A  \fhat ^*   $$
 use
$$ \fhat ^* = S^{-r} \   f^*,  \qquad  \hhat = S^{-r}  \     h    $$

 \noi                 
Multiply  equation (\ref{iJA}) by  $ \sqrt{|\alp |} $ which is given by 
 (\ref{sqalp}).
Remember (\ref{rdeq}),  hence 
$$   {pq \over  4-2p } -2r   =  {q \over  2-p }  $$
Thus we find
$$  i    \sqrt{|\alp |}   J^A = 
S ^{q \over 2-p} \   (f^* \nabo ^A h - h \nabo ^A  f^*) \    
  \sqrt{|\alptil |}                  $$
At this stage it is convenient to notice that  for all scalar function
 $u (x^A)$  we have 
 \beq \nab _1 ^A \   u =  S^{q \over p-2} \   {{\nabtil}_1} ^A   \     u 
                               \label{nabdenab} \eeq
Taking this formula  into account, we can write
$$  i    \sqrt{|\alp |}   J^A =  
(f^*  {\nabtil}_1 ^A h -h {\nabtil}_1 ^A  f^* ) \ 
   \sqrt{|\alptil |}           $$
which proves our assertion, according to (\ref{pGord}).

\subsubsection{Comparison of both methods}
\noi  The first  method is widely employed in the literature concerning 
 FRW universes,  
 and we also used it in the framework of generalized FRW
 spacetimes \cite{LMP}.  As we have just checked, the first method and  the 
second one are equivalent when both can be carried out.
But the former cannot be carried out when $p=2$.
So we are forced to  consider the latter as more fundamental, since
it is not affected by any  dimensional exception. 
In particular it remains  the only one available in the important 
case of a Class $B$ four-dimensional spacetime.  For a  unified approach  that 
encompasses all cases, we are led to a systematic use of the second method.

\section{Sesquilinear Forms for Type I}

In this section we specialize to warped spacetimes of Type I,
with  $V_2$ compact.

\noi  The special case $p=1, q=3$ concerns generalized FRW spacetimes; it has
 been treated in  \cite{LMP} and yields results similar to those of the
 case $p>1$ considered below,   where the quantity defined by formula
  (\ref{sesq1}) is replaced by  $-i W (f^* , h)$ in terms of the Wronskian. 
Extension to arbitrary $q$ is straightforward.

\medskip
\noi   When  $p>1$ we assume in addition that 
        $V_1 \approx {\bf R} \times {\rm compact }$.

\noi        
If $L$ is imbedded in $  (V_1 , \alp) $ as   a   $(p-1)$-dimensional  spacelike 
surface,  let   $dL_A$ be  the    $(p-1)$-dimensional surface  element, 
that is 
\beq   dL_A = {1 \over (p-1)!} \sqrt {|\alp |}  \ 
\varepsilon _{A  B_1 \cdots  B_ {p-1} }  \   dx^{B_1}  \wedge 
         \cdots  dx^{B_ {p-1}}         \label{dLA}  \eeq
Conservation of $J^A$ with respect to $(V_1 , \alp )$ implies that
\beq  (f ; h)_1  =  \int _L   J^A (f, h) \    d L _A    \label {sesq1}   \eeq
doesnot depend on the choice of $L$.
This expression defines a sesquilinear form on $\sron _n$.

\bigskip
\subsection {The Gordon Current}

\noi    Remember usual  formulas  for $N$-dimensional spacetime, with
$\alp , \beta = 0, 1 \dots N-1 $.

\noi   The space of arbitrary solutions to the KG equation is endowed with a 
 sesquilinear form
\beq  (\Phi ; \Omega ) =  \int j^\nu \  d \Sigma _\nu  \label{sesq}  \eeq
$$  j^\nu (\Phi, \Omega) =
-i (\Phi ^*  \nab ^\nu \Omega -    \Omega \nab ^\nu \Phi ^*) $$
Integration is performed over an  $N-1$ dimensional spacelike surface $\Sigma$.
Notation 
 $  \disp d^{N-1} x = dx^1 \wedge  dx^2  \cdots  \wedge dx^{N-1}  $.
 Be cautious that $ d^{N-1}  x $ should not be confused with 
 $d \   x ^{N-1}$.

\noi   Provided  $(\Sigma)$ is  defined by  $x^0 = {\rm const.}  $  we can 
write
 $$  d\Sigma _0 =
{1 \over (N-1) !} \   \eta _{0 1 \cdots  N-1} \
  dx^1 \wedge    \ \cdots    dx^{N-1} = 
   \sqrt {|g|} \   d^ {N-1}  x   $$
For the space components we first have  
$$        d\Sigma _1 = {1 \over (N-1)  !} \  
 \eta _ {1 \alp _1   \cdots     \alp _ {N-1} }
\   dx^\alp _1   \cdots    \wedge  dx^\alp _ {N-1}  $$
Here the indices  $\alp _1 ,  \cdots   \alp _ {N-1}  \not=   1 $,
 thus one of them all 
 must be $0$ (otherwize they would  not be all different, so  
$  \eta _ {1 \alp _1      \cdots   \alp _ {N-1}  }$  would vanish).
Thus $d \Sigma _1$  has  $d x ^0 $ as a factor, say
  $      d\Sigma _1 =  (.....) \wedge  dx^0 $.
But $dx^0$ is zero on  $(\Sigma)$. Thus finally    $  d\Sigma _1 $  vanishes on
 $(\Sigma)$. By a similar  argument  we check that 
      $  d\Sigma _2 $, etc $\cdots  d\Sigma _{N-1} $  also 
vanish on $(\Sig )$.

\noi In order to integrate over  $(\Sigma)$ we evaluate the differential form
$ j \cdot d \Sig $ on this surface; we can write
$$    j \cdot  d \Sigma  =  j^0  d \Sigma  _0                      $$   
$$ j^0 = -i(\Phi ^*  \nab ^0 \Omega -  \Omega  \nab ^0  \Phi ^*)   $$
Naturally $\nab ^0 \Phi = g^{0 \alp }  \partial _ \alp  \Phi $.

    \bigskip 
      \noi
So far the formulas of this subsection are general.
When we   specify to warped spacetimes of Type I, then   $N= p+q$  and   
$V_1$ is  Lorentzian; $V_2$ is Riemannian so  the label  $k$
cannot be $0$. 
In coordinates adapted to the warped product structure, we know that 
$ g ^{Ak} $ all vanish; in particular  $g^{0k}$ vanishes. We are left with 
$$ \nab ^0 \Phi  = g^{0A} \   \partial  _A  \    \Phi                   $$ 
Similarly    
$$ \nab ^0 \Omega  = g^{0A} \   \partial  _A  \    \Omega               $$
            We thus have
\beq   j \cdot  d \Sigma =
-i    \sqrt {|g|} \
d^ {n-1} x \       (\Phi ^*     g^{0A} \   \partial  _A  \    \Omega
-    \Omega          g^{0A} \   \partial  _A  \    \Phi ^*      ) 
                             \label {jdsig}                  \eeq

Cf. (2.2) in ref. \cite{LMP}. 

\noi     Let us now {\em consider  product solutions}.

\noi
 So we assume that $\Phi$ and $\Omega$ are solutions to (\ref{1}) in the form
$$ \Phi =  f(x^A) \    F(x^j)    \qquad    \Omega =  h(x^B) \    H(x^k)     $$
but {\em not necessarily on the same mode}, say
$\Phi \in \hron _n ,  \quad \Omega \in  \hron _l   $
including the  possibility that   $n \not= l$.       We have 
$$  g^{0A} \   \partial  _A  \    \Omega   =   H g^{0A} \partial _A  h $$ 
$$  g^{0A} \   \partial  _A  \    \Phi ^*  =  F^*  g^{0A} \partial _A  f^*  $$
Inserting  into (\ref{jdsig})   yields
$$ j \cdot d \Sigma  = 
-i  \      (\Phi ^* H  g^{0A} \partial _A  h   -    
 \Omega   F^*       g^{0A} \partial _A  f^* ) \      \sqrt {|g|} 
  \        d^{n-1} x        $$
Develop   $ \sqrt {|g|}$   according to  (\ref{det})
where  $S$ and $\gam $ are positive, so 
$$ \disp    \sqrt {|g|}   = 
  \sqrt {|\alp |} \    S ^{q/2}   \       \sqrt \gam        $$
According to  (\ref{defomeg})              we can write
$  \disp        d^{N-1} x =  \omega \wedge  d_2 ^q \  x $, therefore    
$$ j \cdot d \Sigma  =  
-i ( F^*  f^* H      g^{0A} \partial _A  h      -
h H  F^*     g^{0A} \partial _A  f^*)   \sqrt {|\alp |}\
  S ^{q/2}   \               \sqrt \gam
\   \omega   \wedge d _2 ^ q  \  x                       $$
 
$$ j \cdot d \Sigma =
-i ( f^*   g^{0A}    \partial _A   h -  h   g^{0A}    \partial _A f^*)
\  F^*  H        \sqrt {|\alp |} S^{q/2}  \sqrt \gam   
  \           \omega \wedge  d_2 ^q \  x            $$

\noi We now turn to integration.

\noi 
If $p=1$ we can take  
$\Sig = \{ t_0 \} \times V_2$ where $t_0$ is a fixed value of 
the time coordinatr $x^0$. This case 
has been investigated in details in \cite{LMP}. 

\noi         If $p>1$, let us take 
$$  \Sig = L   \times  V_2  \   $$
   where
$ L  \subset  V_1$ is the submanifold defined by $x^0 = {\rm const}$.
Indeed this choice is compatible with the assumption made above that
the equation of $\Sig $ in $V_1 \times V_2$  is just $x^0 = {\rm const}$.

Integrate the above formula; we obtain
\beq  (\Phi ; \Omega ) = 
-i  \left( \int _{V_2}   F^* H \sqrt {\gam}  \    d_2 ^q  x       \right)
\    \int _L  (f^* g^{0A} \partial _A h - h g^{0A} \partial _A f^*)  \  
S^{q/2} \      \sqrt{| \alp |}  \     \omega                  \eeq
\beq      (\Phi ; \Omega ) = 
-i    <F,H>_2                          
\    \int _L  (f^* g^{0A} \partial _A h - h g^{0A} \partial _A f^*)  \  
S^{q/2} \    \sqrt{| \alp |} \    \omega        \label{intform}          \eeq
where we have factorized out the scalar product in $(V_2 , \gam )$
\beq  <F,H>_2 =   \int _{V_2} 
\sqrt { \gam } \    F^*  H  \    d _2 ^q  \   x       \label{scal2}     \eeq
well-defined and positive for arbitrary couple of functions on $V_2$.

 When $F$ and $H$  belong to   ${\cal E} _n$ and
 ${\cal E}_l  $  with $n \not= l$ then   $ <F,H>_2 $ vanishes. 
It follows  that if two product solutions  $\Phi , \   \Omega$
belong to distinct modes, $  ( \Phi ; \Omega) $ vanishes.      Now using 
(\ref{basedev}) and a similar development
$$ \Omega =  \sum  _{s=1} ^{d(l)}  h_s  H_s $$
over a basis $H_1, \dots  $ of  $\eron_l$, we easily check that this property 
holds for any couple of mode solutions, in other words
\begin{prop}
For Type I under our assumptions, two different modes are  orthogonal in 
 the sense of the sesquilinear form defined by the Gordon current.    \end{prop}

[This result extends  Proposition 1 of Ref.\cite{LMP} ]                       

\bigskip                                                   
\noi
When $\Phi$ and $\Omega$ belong to the same mode we give to  (\ref{intform} a 
more compact formulation.               A look at
 (\ref{iJAbis}) (\ref{JdeI}) enables one to write 
\beq j \cdot d \Sigma =    j^0  \cdot  d \Sig _0  =
J^0  \   \sqrt{|\alp |} \   \omega   \    \wedge
( F^* H  \sqrt \gam    \   d_2 ^q \   x )       \label{factor}   \eeq

 We obtain by integration 
$$ \int _\Sig  j \cdot  d \Sig =
<F, H >_2  \   \int _{L}  J^0 \sqrt{|\alp |}  \    \omega         $$
with
At this stage it is convenient to remember that 
 $\omega$ is given by (\ref{defomeg})      
 and to observe that
$$ d L_0 =  \sqrt {|\alp |}  \omega $$
When $A \not= 0$,  if    $  B_1 \cdots  B_ {p-1} $ are all $\not= A $,
one of them must be $0$, thus $d x^0$ is a factor in  $d{L}_A$.

\noi  On $(L)$, we can write $dx^0 = 0$, thus  $dL_A$ vanishes
 for $A \not= 0$.
Thus we have on this manifold
$ J^A  \   d  L _A = J^0 \   d L _0 $. Hence
$$            \int _ \Sig    j \cdot d \Sig =  
<F, H>_2  \   \int _{L}  J^A \   d L _A              $$
A glance at  (\ref{sesq1}) enables us to state
\begin{theo}
In Type I warped spacetime, for product solutions, the sesquilinear map 
defined through the usual Gordon current gets factorized according to the 
formula
$$  ( \Phi ; \Omega ) =
(f ; h ) _1      \    < F , H >  _ 2                                 $$
where  $ (f ; h) _1  $  is defined by (\ref{sesq1}).
     \end{theo}                                                          

\medskip
\section  {Extension to further couplings.}

Up to now we have focussed on the minimal coupling of a free particle to
 gravity.     Extending  our results to  an equation of the form 
\beq  (\nab ^2 + m^2  +  a (x) ) \  \Psi =0     \label{curv} \eeq
is straightforward  {\em provided that the additional term $a$ does not 
depend}    on the $x^j$'s.              Otherwize,
  $\Del _2$ would not commute any more with $ \nab ^2  +  a (x) $, and the 
mode decomposition would be impossible.

\medskip
\noi
Assuming that  $ a = a (x^A) $, 
 the  differential operator in (\ref{curv})  still commutes with
$\Del _2$. Equation (\ref{mod}) is replaced by
$$ (D + \lam S^{-1} + m^2 + a (x^A) ) \  \Phi   =0  $$
Any product solution remains a mode.
In (\ref{fmod})(\ref{rreduc}) $m^2$ must be replaced by
  $m^2 + a(x^A) $  but the  variables $x^j$  remain ignorable.
Since $a$ acts on $f$ as a multiplicative operator, the status of the first
derivatives $\partial f$ in equation (\ref{rreduc}) is not modified;
 therefore condition  (\ref{rdeq}) still avoid the occurence of them. 
Moreover formula (\ref{clef}) which determines $Df$ in terms of
 $\fhat$ is an
 identity valid for arbitrary  $f$.  Again we end up with an equation of the
 form   $\Del _1 \fhat +  \Xi \fhat = 0 $ where $\Xi$ now involves the
 additional term $a$.
   As well-known, this still ensures conservation of $J^A$, given without
    modification by (\ref{Gordhat}).
Finally the results of Section 4 remain valid.

\medskip
\noi    These remarks permit to treat the case where the KG equation includes  
an external potential of the  special (and rather artificial) form $a (x^A)$.
It is more interesting to notice that our study could remain valid for
{\sl curvature  coupling}, characterized by the  addition to  equation
  (\ref{1})       of  a  curvature term
 $ \   \xi  R ^\alp _ \alp \   $ where $\xi$ is a constant
 \footnote{For dimensional reasons, $\xi$ necessarily vanishes in the limit 
where  $\hbar \rightarrow  0$, thus curvature coupling cannot affect the 
geodesic motion},
  and  $ R ^\alp _\alp $ is the scalar curvature of $(V, g)$. 
But the condition which  legitimates this  extension  is that the scalar
curvature of spacetime  depends only on the $x^A$'s.

\noi
We are thus led to investigate what kind of   warped spacetime   supports 
a scalar curvature of the form  $R (x^A)$.

The following lemma will be useful.
\begin{lem} 
Let $u(x^A)$ be a function on $V$ satisfying  $\partial _j u = 0$.
Then the following quantities
$$ \nab ^A \partial _B u  ,   \      \qquad
\nab ^A  u  ,  \       \qquad
\nab ^A \nab _A   \    u                                    $$
are independent from the coordinates $x^k$.
\end{lem}

\noi Proof:

\noi  $\nab _A \partial _B  \  u  = 
\partial _A \partial _B  u   - {\Gam _A ^{\   \mu} }_B \partial _\mu \   u  $.
The first term is obviously independent from $x^k$'s. 
Since  $\partial _ j u $ is zero, 
 only the coefficients   $ {\Gam _A ^{\    C} }_B  $    give a contribution
to the second term. But it is known that
      $ {\Gam _A ^{\    C} }_B  =      \  ^1 {\Gam _A ^{\    C} }_B        $,
where    $ \   ^1 \Gam  $ is the connexion for $(V_1 , \alp)$  (see 
\cite{oneil}; in 4 dimensions see formula (26) in \cite{carot}.
Thus the second term has the same property.

\noi The statement about  $ \nab ^A  u $ is obvious, if we remember that
$g ^\AB = \alp ^\AB,    \    g^{A0}  =0$.

\noi Finally, we have    
$ \nab ^A \nab _A   \    u  =   \alp ^\AB  \nab _B   \partial _A  \   u $
which is independent of $x^k$ because of the first statement.

\bigskip
 \noi  The Ricci tensor of a warped product can be expressed in terms of
 the warping function and of the geometry of the factor manifolds 
(See O'Neill \cite{oneil}, especially   corollary 43, p. 211.
When  $p+q=4$,  see Carot and da Costa  
\cite{carot}, eqs (25),  for a more transparent notation).   

We are concerned with
\beq R =  \alp ^\AB  R_ \AB 
  - S^{-1} \gam ^\ij R_\ij                 \label{ricc}     \eeq
In the first term, we know that  
$ R _\AB $  differs from  $  \    ^1  R_\AB $  only by  a function of  
$ \disp e ^\Theta $   and $ \disp   \nab _B  \nab _A  (e^\Theta) $, which 
cannot depend on $x^C$ according to the previous result. 
Therefore   $R_\AB $ doesnot depend on the  $x^k$'s 
$$ \partial   _j  R _\AB   = 0                         $$
This result, in turn,  entails that  $R$ might depend on these  variables 
only through   the quantity   $\gam ^\ij   R_ \ij$. 

\noi According to the literature  cited,  $R_\ij $  takes on the form
$$ R_\ij =   \   ^2 R_\ij   +  N  \gam _\ij                      $$
where    $N$ is a function    of  
  $\Theta , \    \nab ^A \Theta , \    \partial _A \Theta  $and 
$\nab ^A   \nab_A   \   \Theta $.     By the above lemma, we see that
$$\partial _ j  N = 0$$
By contraction we obtain
$$\gam ^\ij   R_ \ij =    \gam ^\ij   \   \  ^2 R_\ij    +  q N            $$
It is now clear that   $\gam ^\ij   R_ \ij $,  and therefore equally
$R$, depends on the $x^k$'s except when $\   ^ 2  R$ is a constant; thus

\begin{theo}
(Only) if $(V_2 , \gam ) $ has constant scalar curvature  
$  \gam ^\ij  \   ^2 R_\ij $,
 the scalar curvature $R$ 
of the warped product can be regarded as a function on $V_1$.
\end{theo}

An well-known instance of this situation is the case of 
conventional FRW spacetime,
where the space sections are homogeneous. 

 \section{Concluding remarks}
We have separated the variables in the KG equation with help of  a generalized
mode decomposition which is  possible  essentially because the motion of a 
test particle in a warped product spacetime admits a remarkable first 
integral.

\noi At least for Type I, these modes are actually "normal" , that is 
orthogonal, in the sense of the usual sesquilinear form associated with the 
Gordon current. This form itself has been analysed in terms of the vector 
field  $J$  defined on the Lorentzian factor manifold and, under very large 
assumptions, a  sesquilinear form has been defined on the solutions to the 
{\em reduced} wave equation.

\noi Type I is  perhaps physically the most interesting; in 4 dimensions it 
encompasses not only generalized FRW spacetimes, but also all kind of 
spherically symmetric spacetime,  including the nonstationary ones.
However further investigations of Type II might be of interest.

\noi It is noteworthy that the whole picture remains valid in the presence of a 
curvature coupling term $\xi R^\alp _\alp $ only under the condition that
$(V_2 , \gam )$ has  a constant scalar curvature. 
For FRW, abandoning spatial homogenity would not permit to carry out our mode 
decomposition when a curvature coupling term is introduced into the KG 
equation. In contrast, the mode decomposition remains  possible with such a 
term for all kind of spherically symmetric spacetime.

\end{document}